\documentclass[aps,showpacs,showkeys]{revtex4}
\usepackage{graphics,epsfig,psfig}

\begin{document}

\def\be{\begin{equation}}
\def\ee{\end{equation}}
\def\bea{\begin{eqnarray}}
\def\eea{\end{eqnarray}}
\def\ket#1{\hbox{$\vert #1\rangle$}}   
\def\bra#1{\hbox{$\langle #1\vert$}}   
\def\oneh{{\textstyle {1\over 2}}}
\def\onet{{\textstyle {1\over 3}}}
\def\smoneh{{\scriptstyle {1\over 2}}}
\def\onesix{{\textstyle {1\over 6}}}
\def\oneq{{\textstyle {1\over 4}}}
\def\treh{{\textstyle {3\over 2}}}
\def\treq{{\textstyle {3\over 4}}}
\def\oneight{{\textstyle {1\over 8}}}
\def\onesq{{\textstyle {1\over \sqrt{2}}}}

\def\Re{\hbox{\rm Re\,}}
\def\Im{\hbox{\rm Im\,}}

\def\Tr{\hbox{\rm Tr\,}}
\def\Sp{\hbox{\rm Sp\,}}
\def          
\simeq{{\ \lower2pt\hbox{$-$}\mkern-13mu \raise2pt \hbox{$\sim$}\ }} 
\def\dirac#1{\slash \mkern-10mu #1}

\title{Nonperturbative versus perturbative effects in generalized parton
distributions}

\author{B.~Pasquini$^{a,b}$, M.~Traini$^{b,c}$, S.~Boffi$^a$}

\affiliation{
$^a$Dipartimento di Fisica Nucleare e Teorica, Universit\`a degli
Studi di Pavia and INFN, Sezione di Pavia, Pavia, Italy}
\affiliation{
$^b$ECT$^*$, Villazzano (Trento), Italy}
\affiliation{
$^c$Dipartimento di Fisica, Universit\`a degli Studi di Trento, Povo
(Trento), and INFN, Gruppo Collegato di Trento, Trento, Italy}

\begin{abstract}
Generalized parton distributions (GPDs) are studied at the hadronic
(nonperturbative) scale within  different assumptions based on a relativistic
constituent quark model. In particular, by means of a meson-cloud model we
investigate the role of nonperturbative antiquark degrees of freedom and the 
valence quark contribution. A QCD evolution of the obtained GPDs is used to 
add perturbative effects and to investigate the GPDs' sensitivity 
to the nonperturbative ingredients of the calculation at larger 
(experimental) scale.
\end{abstract}


\maketitle


\section{Introduction}

Generalized parton distributions (GPDs) are basic ingredients in the description
of hard exclusive processes (see ref.~\cite{diehl} and references therein). Not
only are they generalizations of the well known parton distributions from
inclusive deep inelastic scattering (DIS), but being correlation functions they
also incorporate non-trivial behavior of hadrons related to the nonperturbative
regime of quantum chromodymanics (QCD). At present, apart from the preliminary
studies of lattice QCD~\cite{negele,goeckeler}, one cannot calculate GPDs
from first principles, so one has to resort to models or parametrizations. 

In fact, the perturbative approach to QCD is able to connect observables at
different resolution scales, but the realization of the complete project (i.e.
to fully understand the consequences of the dynamics of quarks and gluons)
requires the input of unknown nonperturbative matrix elements to provide
absolute values for the observables at any scale. In the present paper we intend
to apply a radiative parton model procedure which, starting from a low
resolution scale $Q_0^2$, has been able to reproduce and predict (see, e.g.,
ref.~\cite{RPM} and references therein) the main features of the 
experimental deep inelastic
structure functions at high momentum transfer. The procedure assumes that there
exists a scale where the short range (perturbative) part of the interaction is
negligible and, therefore, the glue and sea are suppressed, and a long range
(confining) part of the interaction produces a proton composed by (three)
valence quarks, mainly~\cite{PaPe}. Jaffe and Ross~\cite{JaRo} proposed to
ascribe the quark model calculations of matrix elements to that hadronic scale
$Q_0^2$. In this way, quark models summarizing a great deal of hadronic
properties may substitute for low-energy parametrizations, while evolution to
larger $Q^2$ is dictated by perturbative QCD.

In the following we study the nucleon's GPDs within specific hadron models
and address the problem of evolving the input distributions to the experimental
scale investigating the effects of different dynamical assumptions. In 
particular, we want to investigate both the quark core structure of the nucleon 
and its chiral properties. 
In fact, the new aspects of the GPDs with respect to the better known parton 
distributions are related to the socalled ERBL region where the presence of
dynamical $q\bar q$ pairs, both in the nonperturbative and perturbative regimes,
plays a crucial role. 

The importance of the chiral structure of nucleons is well established both
experimentally and theoretically. The pion cloud associated with chiral symmetry
breaking was first discussed in the DIS context by
Feynman~\cite{FeyPHI} and Sullivan~\cite{Sull72}. It leads to flavor symmetry
violations in the sea-quark distribution of the nucleons~\cite{Thomas83},
naturally accounting for the excess of $\bar d$ (anti)quarks over $\bar u$
(anti)quarks as observed experimentally through the violation of the Gottfried
sum rule~\cite{NMC,NA51,E866,Hermes}. As discussed by Melnitchouk {\it et
al.}~\cite{MST99}, the relatively large asymmetry found in these experiments
implies the presence of nontrivial dynamics in the proton sea which does not
have a perturbative QCD origin. In particular, a quantitative description of the
entire region of the quark momentum fraction $x$ covered by the experiments 
requires a delicate balance between several competing mechanisms. At larger $x$
the dynamics of the pion cloud provides the bulk of the $\bar d -\bar u$
asymmetry with DIS from the $\pi N$ component of the
nucleon wave function, however also the $\pi \Delta$ arises in the light-cone
formulation of the meson-cloud model and it is of some importance in
DIS too.

Although the nucleon's nonperturbative antiquark sea cannot be attributed
entirely to its virtual meson cloud~\cite{Koepf}, the role of mesons in DIS is
of primary importance, and the idea was developed further giving origin to the
meson-cloud model (for a review of early work, see refs.~\cite{STh98,Londergan,
Kumano} and references therein). The connection between this model and the
chiral properties of QCD was established by investigating the
nonanalytic behavior of the $\bar d -\bar u$ distribution~\cite{ThMS00} (see,
however, \cite{chen-ji}). The meson-cloud picture is also suggested by QCD in 
the limit of large numbers of colors $N_c$, where it becomes equivalent to an
effective theory of mesons, in which baryons appear as solitons, i.e. classical
solutions characterized by a mean meson field~\cite{Witten}. A realization of
this idea is achieved in the chiral quark-soliton model, where the effective
action is derived from the instanton vacuum of QCD, thus providing a microscopic
mechanism for the dynamical breaking of chiral symmetry~\cite{Diakonov}. Flavor
asymmetry of the antiquark distributions arises in this model of the nucleon as
a sub-leading effect in the limit of large $N_c$~\cite{Dressler}.

An alternative approach to investigate the role of $q\bar q$-pairs in DIS and to
access the ERBL region, is considering the constituent quarks as complex
systems~\cite{ACMP}. Such a scheme has been recently developed in relation to a
nonrelativistic constituent quark model, both for parton
distributions~\cite{SVT} and GPDs~\cite{SV}. 

In the present work we will study the possibility of integrating meson-cloud
model effects into the evolution of GPDs. To this end we will assume that GPDs
can be written in terms of double distributions~\cite{Rad99}, involving a given
profile function and the forward parton distribution derived in some model. At
the same time the model proposed by Melnitchouk {\it et al.}~\cite{MST99} will
be adapted to show how the knowledge of the meson-cloud effects can be
incorporated within a relativistic quark-valence approach to GPDs. The double 
distribution (DD) model will be briefly recalled in sect.~\ref{sect:GPDs-DD}.
Here the input parton distribution is discussed both in terms of the pure
valence contribution derived in light-front relativistic quark models
(sect.~\ref{sect:LFCQM}) and in the meson-cloud model
(sect.~\ref{sect:fromdiagonal}). Matching sea, gluons and valence-parton
distributions in QCD evolution of the obtained GPDs is then briefly described in
sect.~\ref{sect:Match}, and the results are discussed in sect.~\ref{sect:RES}.
Some conclusions are drawn in the final section. 


\section{Modeling GPDs with double distributions}
\label{sect:GPDs-DD}

In the following we shall concentrate our attention on the chiral even (helicity
conserving) distribution $H^q(\overline x,\xi,Q^2,t)$ for partons of flavor $q$
at the hadronic scale where the models we are going to discuss are 
assumed to be valid to evaluate the twist-two amplitude. Such amplitude occurs, 
for example, in deeply virtual Compton scattering where a lepton exchanges a 
virtual photon of momentum $q^\mu$ with a nucleon of
momentum $P^\mu$, producing a real photon of momentum ${q'}^\mu$ and a recoil
nucleon of momentum ${P'}^\mu$. Then
$Q^2 = -q^\mu q_\mu$ is the space-like virtuality that defines the scale
of the process.
In a symmetric frame of reference where $q^\mu$
and the average nucleon momentum $\overline P^\mu = \oneh(P^\mu + {P'}^\mu)$ are
collinear along the $z$ axis and in opposite directions, 
the quark light-cone momentum is $\overline k^+ = \overline x \overline P^+$,
the invariant momentum square is $t=\Delta^2=({P'}^\mu-P^\mu)^2$, and the
skewedness $\xi$ describes the longitudinal change of the nucleon momentum,
$2\xi=-\Delta^+/\overline P^+$. 

For sake of simplicity we follow the common notation which explicitly indicates 
three variables only $(\overline x,\xi,t)$ instead of four $(\overline
x,\xi,Q^2,t)$. We shall come back to the definition of $Q^2$ when we discuss the
values of the hadronic scale $Q_0^2$ and the QCD evolution of $H^q$ in $Q^2$.

We also introduce non-singlet (valence) and singlet quark distributions,
\be  
\label{eq:a12}   
H^{NS} (\overline x, \xi,t)  \equiv  
\sum_q\left [H^q (\overline x, \xi,t)  +  H^q (-\overline x, \xi,t) \right ] 
=  H^{NS} (-\overline x, \xi,t), 
\ee
\be 
H^S (\overline x, \xi,t)  \equiv  
\sum_q \left [H^q (\overline x, \xi,t) -  H^q (-\overline x, \xi,t)  \right ] 
=  - H^S (-\overline x, \xi,t),  
\ee 
respectively. Besides being symmetric or antisymmetric in $\overline x$, they
are also symmetric under $\xi\to-\xi$ due to the polynomiality
property~\cite{jig}.

The analogous GPD for gluons is symmetric in $\overline x$, i.e.
\be
H^g(\overline x,\xi,t) = H^g(-\overline x,\xi,t) ,
\ee
and reduces to the gluon density $g(x)$ in the forward limit ($\overline x\to
x$) 
\be
H^g(x,0,0) = x\,g(x), \quad x>0.
\ee

There are two distinct regions: the Dokshitzer-Gribov-Lipatov-Altarelli-Parisi
(DGLAP) region, $\vert\overline x\vert > \xi$, and the
Efremov-Radyushkin-Brodsky-Lepage (ERBL) region, $\vert\overline
x\vert < \xi$. The naming derives from the fact that the GPD evolution
equations in the region $\vert\overline x\vert>\xi$ ($\vert\overline x\vert<\xi$)
reduce to the familiar DGLAP (ERBL) equations in the limit $\xi=0$ ($\xi=1$).

The singlet and gluon distributions mix under evolution, while the non-singlet
distribution does not. 

GPDs depend on the invariant momentum transfer $t$. In particular, the first
moment in $\overline x$ of $H^q(\overline x,\xi,t)$ is independent of $\xi$ and
related to the Dirac form factor of the proton. Thus in phenomenological
constructions of GPDs it has been found convenient to assume a factorized 
$t$ dependence determined by some form factors. This simplifies the QCD evolution
considerably because in this way the $t$ dependences of quarks and gluons (which
mix under evolution) are not modified during evolution.

The $t$-independent part $H^q(\overline x, \xi) \equiv H^q(\overline x, \xi, t =
0)$ is parametrized by a two-component form~\cite{Rad99}:
\be
H^q(\overline x, \xi) = H^q_{DD}(\overline x, \xi) +
\theta(\xi-|\overline x|)\, 
 D^q\left(\frac{\overline x}{\xi}\right) ,
\label{eq:dd}
\ee
where 
\be
H^q_{DD}(\overline x,\xi)= \int_{-1}^{1}d\beta
\int_{-1+|\beta|}^{1-|\beta|} d\alpha\,
\delta(\overline x-\beta-\alpha\xi)\, F^q(\beta,\alpha) .
\label{eq:dd2}
\ee
The D-term contribution $D^q$ in eq.~(\ref{eq:dd}) completes the parametrization
of GPDs, restoring the correct polynomiality properties of
GPDs~\cite{jig,Pol99b}. It has a support only for $|\overline x| \leq \xi$, so
that it is invisible in the forward limit. The D-term contributes to the
singlet-quark and gluon distributions and not to the non-singlet distribution.
Its effect under evolution is at the level of a few percent~\cite{FmcD02b}, and
in the following it will be disregarded in the input GPDs.

According to Radyushkin's suggestion~\cite{Rad99}, the DDs 
entering eq.~(\ref{eq:dd2}) are written as
\be
F^q(\beta, \alpha) = h(\beta, \alpha) \, q(\beta) ,
\label{eq:ddunpol}
\ee
where $q(\beta)$ is the forward quark distribution (for the flavor
$q$), and the profile function $h(\beta, \alpha)$ is parametrized
as~\cite{Rad99,Rad01b}
\begin{eqnarray}
h(\beta , \alpha) = 
 \frac{\Gamma(2b+2)}{2^{2b+1}\Gamma^2(b+1)}\,
\frac{\bigl[(1-|\beta|)^2-\alpha^2\bigr]^{b}}{(1-|\beta|)^{2b+1}} .
\label{eq:profile}
\end{eqnarray}
In eq.~(\ref{eq:profile}), the parameter $b$ determines the width of the profile
function $h(\beta , \alpha)$ and characterizes the strength of the
$\xi$ dependence of the GPDs. It is a free parameter for the valence 
($b_{val}$)
and sea ($b_{sea}$) contributions to GPDs, which can be used in such
an approach as fit parameters in the extraction of GPDs from hard
electroproduction observables. The favoured choice is 
$b_{val} = b_{sea} = 1.0$,
corresponding to maximum skewedness. With a similar assumption adopted for the
gluon distribution one defines $b_{gluon}=2$. The limiting case $b\to\infty$
gives $h(\beta, \alpha)\to\delta(\alpha)\,h(\beta)$ and corresponds to the
$\xi$-independent ansatz for the GPD, i.e. $H^q(\overline x,\xi)\to
H^q(x,\xi=0)=q(x)$, as used in refs.~\cite{Gui98,Vdh98}. 

In order to explicitly calculate $H^q(\overline x, \xi)$ in eq.~(\ref{eq:dd})
knowledge of the parton distribution $q(x)$ is needed. In the two following
subsections details are given about the derivation of $q(x)$.


\subsection{Parton distributions and light-front relativistic quark models}
\label{sect:LFCQM}

Following the lines of ref.~\cite{diehl2} in two recent
papers~\cite{BPT03,BPT04} we discussed a method to evaluate 
GPDs within light-front constituent quark models (CQMs) at the scale dominated
by valence (constituent) degrees of freedom. 
The comparison of these calculations with predictions  in
the chiral soliton model and the MIT bag model, as well as the consistency  
with lattice results for the first moments of GPDs 
showed that all the phenomelogy for large $\bar x$ and small $t$ 
could be studied within the assumed relativistic CQM.
As a drawback of such an approach, the calculation was
restricted to the region ${\overline x} \geq \xi$ and the generation of $q\bar
q$ contributions could have a perturbative origin only. In contrast, within the
DD-based model both ERBL and 
DGLAP regions of quark GPDs are populated with any input quark distribution
$q(x)$ (with or without sea contribution). In this subsection 
we only consider valence quarks leaving for the next subsection the case
including the sea.

According to the approach of refs.~\cite{BPT03,BPT04} 
the parton distribution in
relativistic light-front CQM takes this simple form: 
\be
\label{eq:q-val}
q(x,\mu_0^2) = \sum_{j=1}^3\delta_{\tau_j\tau_q}\int
\prod_{i=1}^3 d\vec{k}_i \, \delta\left(\sum_{i=1}^3 \vec{k}_i\right)
\,\delta\left(x-\frac{k^+_j}{M_0}\right)
\vert\Psi_\lambda^{[c]}(\{\vec{k}_i;\lambda_i,\tau_i\})\vert^2,
\ee
where $k_j^+=(k^0_j+k_j^3)/\sqrt{2}$ is
the quark light-cone momentum, and $M_0=\sum_i\, \sqrt{{\vec k}^2_i+m^2_i}$ is 
the free mass for the three-quark system. 
$\Psi_\lambda^{[c]}(\{\vec{k}_i;\lambda_i,\tau_i\})$ is the canonical wave
function of the nucleon in the instant form; under the assumption that only
valence quarks are active, it is obtained by solving an eigenvalue equation for
the mass operator within relativistic CQMs. 

In the following we will discuss results based on the mass operator for the
hypercentral CQM~\cite{FTV99}, i.e.
\be
\label{eq:hypmass}
M = \sum_{i=1}^3 \sqrt{\vec{k}_i^2 + m_i^2} -\frac{\tau}{y} +\kappa_l \,y,
\ee
with $\sum_i\vec{k}_i = 0$, and $m_i$ being the constituent quark masses,
$y=\sqrt{\vec{\rho}^2 + \vec{\lambda}^2}$ is the radius of the
hypersphere in six dimensions and $\vec{\rho}$ and $\vec{\lambda}$ are 
Jacobi coordinates. For a discussion of the model see refs.~\cite{FTV99,GE95}.

The distribution (\ref{eq:q-val})
automatically fulfills the support condition and satisfies the (particle)
baryon number and momentum sum rules at the hadronic scale $\mu_0^2$ where the
valence contribution dominates the twist-two response: 
\be
\int dx \,q(x,\mu_0^2) = N_q,
\ee
with $N_q$ being the number of valence quarks of flavor $q$,
\be
\int dx\, x\, [u(x,\mu_0^2) + d(x,\mu_0^2)] = 1,
\ee
and $u(x,\mu_0^2)\equiv u_V(x,\mu_0^2)$ and 
$d(x,\mu_0^2)\equiv d_V(x,\mu_0^2)$, 
the up and down valence-quark distributions.


\subsection{Parton distributions and the meson-cloud model}
\label{sect:fromdiagonal}

Let us now introduce the meson-cloud model to incorporate $q \bar q$
contributions into the valence-quark model of the parton distribution discussed
in the previous section.

The basic hypothesis of the meson-cloud model is that the physical nucleon state
can be expanded (in the infinite momentum frame (IMF) and in the one-meson
approximation) in a series involving bare nucleons and two-particle,
meson-baryon states. Its wave function is written as the sum of meson-baryon
Fock states
\begin{equation} 
   |p \rangle = \sqrt{Z}|p\rangle_{bare} 
+\sum_{BM}\int dy\, d^2\vec{k}_\perp\, \phi_{BM}(y,k_\perp^2) \,
 |B(y,\vec{k}_\perp );M(1-y,-\vec{k}_\perp ) \rangle . 
\label{eq:dressed}
\end{equation} 
Here $\phi_{BM}(y,k_\perp^2)$ is the probability amplitude for the proton to
fluctuate into a virtual baryon-meson $BM$ system with the baryon and meson
having longitudinal momentum fractions $y$ and $1-y$ and transverse momenta 
$\vec{k}_\perp$ and $-\vec{k}_\perp$, respectively. $Z$ is the wave function
renormalization constant and is equal to the probability to find the bare proton
in the physical proton. 

The lowest lying fluctuations for the proton which we include in our calculation
are
\begin{eqnarray} 
     p (uud) &\rightarrow  & n(udd) \,\pi^+(u\bar d),\nonumber\\ 
     p (uud) &\rightarrow  & p(uud)\,  
 \pi^0 \left( \frac{1}{\sqrt{2}} [d\bar d - u\bar u]\right),\nonumber\\ 
     p (uud) &\rightarrow  & \Delta^+(uud) 
 \,\pi^0 \left(\frac{1}{\sqrt{2}} [d\bar d - u\bar u]\right),\nonumber\\ 
  p (uud) &\rightarrow  & \Delta^{0} (udd) \,\pi^+(u \bar d),\nonumber\\ 
  p (uud) &\rightarrow  & \Delta^{++} (uuu) \,\pi^-(\bar u d).   
\end{eqnarray}     
In DIS the virtual photon can hit either the bare proton
$p$ or one of the constituents of the higher Fock states. In the IMF, where the
constituents of the target can be regarded as free during the interaction time,
the contribution of the higher Fock states to the quark distribution of the
physical proton, can be written as the convolution
\begin{equation}
 \delta q_{p}(x) = \sum_{MB} \left [
  \int_x^1 \frac{dy}{y}\,f_{MB/p}(y)\, q_M \left(\frac{x}{y}\right)
 +  \int_x^1 \frac{dy}{y}\, f_{BM/p}(y)\, q_B \left(\frac{x}{y}\right)
 \right ] ,
\label{eq:splitting}
\end{equation}
where the splitting functions $f_{MB/p}(y)$ and $f_{BM/p}(y)$ are related to the
probability amplitudes $\phi_{BM}$ by 
\begin{equation}
f_{BM/p}(y) = f_{MB/p}(1-y) 
= \int_0^\infty dk_\perp^2 \sum_{\lambda,\lambda^\prime} 
\left\vert\phi_{BM}^{\lambda\lambda^\prime} (y,k_\perp^2)\right\vert^2. 
\end{equation}
Here $\phi_{BM}^{\lambda\lambda^\prime} (y,k_\perp^2 )$ is the probability 
amplitude for a hadron with given positive helicity  to be in a Fock state
consisting of a baryon with helicity $\lambda$ and a meson with helicity
$\lambda^\prime$~\cite{BoTho99}. It can be calculated by using time-ordered
perturbation theory in the IMF. The quark distributions in a physical proton are
then given by
\begin{equation}
q_{p} (x) = Z q_{p}^{bare}(x) + \delta q_{p}(x) , 
\label{eq:partondis}
\end{equation}
where $q_{p}^{bare}$ are the bare quark distributions and the renormalization
constant $Z$ is given by 
\begin{equation}
Z\equiv 1 - \sum_{MB} \int_0^1  dy\,f_{MB/p}(y).
\label{eq:renorm}
\end{equation}
The amplitudes $\phi^{\lambda\lambda^\prime}_{BM}(y,k_\perp^2)$ may be expressed
in the following form 
\begin{equation}
\phi^{\lambda\lambda^\prime}_{BM}(y,k_\perp^2)
 = \frac{1}{2\pi\sqrt{y(1-y)}} 
  \frac{\sqrt{m_H m_B}} {m^2_H-{\cal M}_{BM}^2(y,k_\perp^2 )}
  \,G_{HBM}(y,k_\perp^2 ) V^{\lambda\lambda^\prime}_{IMF}(y,k_\perp^2), 
\end{equation}
where $m_H$ is the physical mass of the fluctuating hadron (in present case 
the proton, but the approach can be generalized (e.g. ref.~\cite{BoTho99})), and 
\begin{equation}
{\cal  M}_{BM}^2 =\frac{k_\perp^2 + m_B^2}{y} +
\frac{k_\perp^2 + m_M^2}{1-y}
\end{equation}
is the invariant mass of the meson-baryon fluctuation.
$V^{\lambda\lambda^\prime}_{IMF}(y,k_\perp^2)$ describes the vertex and 
contains  the spin-dependence of the amplitude. The exact form of the
$V^{\lambda\lambda^\prime}_{IMF}(y,k_\perp^2)$ can be found for various
transitions in refs.~\cite{STh98,HSS96}. Because of the extended nature of the
vertices one has to introduce phenomenological vertex form factors, 
$G_{HBM}(y,k^2_\perp )$, which parametrize the unknown dynamics at the vertices.
We use the popular parametrization
\begin{equation} 
     G_{HBM}(y,k_\perp^2 )= 
\left( \frac{\Lambda^2_{BM}+ m_H^2}{
\Lambda^2_{BM}+{\cal M}^2_{BM}(y,k_\perp^2)}\right)^2 . 
\end{equation}

In order to calculate the meson-cloud corrections to the quark distributions we
have to specify the coupling constants entering
$V^{\lambda\lambda^\prime}_{IMF}(y,k_\perp^2)$ and the cut-off parameters
$\Lambda_{BM}$. We use the numerical values as given by~\cite{MST99,BoTho99},
i.e. $g^2_{NN\pi}/4\pi=13.6$ and $g^2_{N\Delta\pi}/4\pi = 11.08$ GeV$^{-2}$. 
The couplings of a given type of fluctuation with different isospin components
are related by isospin Clebsch-Gordon coefficients, 
$g_{pn\pi^+}=- \sqrt{2}g_{pp\pi^0}$, 
$g_{p\Delta^0\pi^+}
= -g_{p\Delta^+\pi^0}/\sqrt{2}
= g_{p\Delta^{++}\pi^-}/\sqrt{3}$, 
with $g_{NN\pi}=g_{pp\pi^0}$ and $g_{N\Delta\pi}=g_{p\Delta^{++}\pi^-}$.
The violation of the Gottfried sum rule and flavor symmetry puts also
constraints on the magnitude of the cut-off parameters. The values
$\Lambda_{MB} =1.0$ GeV and $\Lambda_{MB} =1.3$ GeV for the $\pi N$ and
$\pi\Delta$ components, respectively, give contributions to the $\bar u$ and
$\bar d$ which are consistent with the requirement that the
meson-cloud component of the sea quark contribution 
cannot be larger than the measured sea
quark and also with flavor symmetry violation~\cite{MST99}. With this choice of
the parameters the probabilities of the fluctuations are given by 
$P_{N\pi/p} = 13\%$, $P_{\Delta\pi/p} = 11\%$.

In the following we will assume that at the lowest hadronic scale the bare
nucleon is described by the relativistic quark model wave function formulated
within the light-front dynamics and, as a consequence, that only valence partons
will contribute to the partonic content of the bare nucleon~\cite{FTV99,BPT03}.
The full (nonperturbative) antiquark content will be generated by the
meson-cloud mechanism described by eq.~(\ref{eq:splitting}). The partonic  
content of the $\Delta$ and the pion will be consistently evaluated
within the same scheme assuming light-front dynamics and valence contributions 
only. Within these approximations the meson-cloud corrections at the hadronic
scale $\mu_0^2$ read
\begin{eqnarray}
  q_{p}(x) = Z q_p^{bare}(x) 
&+& \int_x^1  \frac{dy}{y}\left[f_{N\pi/p}(1-y) + f_{\Delta\pi/p}(1-y)
  \right] q_\pi \left(\frac{x}{y}\right)  \nonumber\\
&+& \int_x^1  \frac{dy}{y}\left[f_{N\pi/p}(y) + f_{\Delta\pi/p}(y)
  \right]  q_\Delta \left(\frac{x}{y}\right),
\label{eq:qp-cloud}
\end{eqnarray}
where $q_{p} \equiv (u=u_V+\bar u,d=d_V+\bar d)$ include both valence and sea
contribution coming from the meson-baryon fluctuations, while 
$q_{p}^{bare}\equiv (u_V^{bare}, d_V^{bare})$ include the valence contribution
only as discussed in section~\ref{sect:LFCQM}. 
The conservation of both momentum
and baryon number sum rules is guaranteed by the correct formulation of
meson-cloud approach, in particular by the momentum conservation due to the
symmetry $f_{BM/p}(y) = f_{MB/p}(1-y)$ in eq. (\ref{eq:splitting}) and by the
renormalization factor $Z$ of eqs. (\ref{eq:dressed}), (\ref{eq:partondis}) and
(\ref{eq:renorm}).

\hspace{0.5cm}

\noindent {\it {\bf The nucleon}}

\hspace{0.5cm}

In order to model the partonic content at the scale $\mu_0^2$ for the nucleon,
the $\Delta$ and the pion, we make use of the light-front approach
discussed in sect.~\ref{sect:LFCQM} and calculate the diagonal component 
of the generalized parton distributions, i.e. the inclusive parton distributions
by means of
\be
q_p^{bare}(x) \equiv q(x,\mu_0^2),
\ee
where $q(x,\mu_0^2)$ is given by eq.~(\ref{eq:q-val}).

\hspace{0.5cm}

\noindent {\it {\bf The $\Delta$}}

\hspace{0.5cm}

The calculation of the cloud contribution involves the explicit form for the
parton distributions $q_\Delta (x)$ of the $\Delta$ (see eq.
(\ref{eq:qp-cloud})); we use the results of the relativistic model
for the nucleon and the isospin symmetries:
\be
\begin{array}{lcllcl}
u_{\Delta^{++}}(x,\mu_0^2) &= &\frac{3}{2} u_{p}(x,\mu_0^2), \quad
&d_{\Delta^{++}}(x,\mu_0^2) &= &0,                             \\
u_{\Delta^{+}}(x,\mu_0^2)  &= &u_{p}(x,\mu_0^2), \quad
&d_{\Delta^{+}}(x,\mu_0^2)  &= &d_{p}(x,\mu_0^2),              \\
u_{\Delta^{0}}(x,\mu_0^2)  &= &\frac{1}{2}u_{p}(x,\mu_0^2), \quad
&d_{\Delta^{0}}(x,\mu_0^2)  &= &2\,d_{p}(x,\mu_0^2),           \\
u_{\Delta^{-}}(x,\mu_0^2)  &= &0, \quad
&d_{\Delta^{-}}(x,\mu_0^2)  &= &3\,d_{p}(x,\mu_0^2).          
\end{array}
\ee

\hspace{0.5cm}

\noindent{\it {\bf The pion}}

\hspace{0.5cm}

The canonical wave function of the pion is taken from ref.~\cite{Choi99} and
reads 
\begin{eqnarray}
\Psi^{[c]}(\vec{k}_1,\vec{k}_2;\mu_1,\mu_2)
=\frac{1}{\pi^{3/4}\beta^{3/2}}
\left(\oneh\mu_1\oneh\mu_2|00\right) \exp{(-k^2/(2\beta^2))},
\label{eq:can_psi}
\end{eqnarray}
with $\vec k=\vec k_1=-\vec k_2$, $x=x_1=k^+/M_0,$ $x_2=1-x$,  
$M_0^2=(\vec k_\perp^2+m_q^2)/x+(\vec k_\perp^2+m_q^2)/(1-x)$, and
$\beta=0.3659$ GeV.
The choice of the model from ref.~\cite{Choi99} is consistent with the
hypercentral CQM we adopt for the nucleon, in fact the central potential between
the two constituent quarks is described as a linear confining term plus
Coulomb-like interaction. The canonical expression (\ref{eq:can_psi}) represents
a variational solution to the mass equation.

The light-front parton distribution of the $\pi^+$ is given by
\begin{eqnarray}
q_{\pi^+}(x) = v_\pi(x,\mu_0^2) &=& \sum_{j=1}^2\delta_{\tau_j\tau_q}\int
\prod_{i=1}^2 d\vec{k}_i \, \delta\left(\sum_{i=1}^2 \vec{k}_i\right)
\,\delta\left(x-\frac{k^+_j}{M_0}\right)
\nonumber\\
& &\qquad\times 
\left\vert\Psi_\lambda^{[c]}(\{\vec{k}_i;\lambda_i\})\right\vert^2
\left\vert\left(\oneh\tau_j\oneh-\tau_j|00 \right)\right\vert^2.
\end{eqnarray}
Isospin symmetry imposes 
$u_V^{\pi^+}=\bar d_V^{\pi^+}=\bar u_V^{\pi^-}=d_V^{\pi^-}=v_\pi(x,\mu_0^2)$,
while, due to the model restrictions, the pion sea at the hadronic scale
vanishes: $\bar u^{\pi^+}= d^{\pi^+}=u^{\pi^-}=\bar d^{\pi^-}=0$.


\section{Matching sea, gluons and valence-parton distributions in QCD evolution}
\label{sect:Match}

In order to extract the parton distributions of the nucleon including
the sea (cloud) contributions we need to match sea, valence and gluons within
the radiative parton model and to identify the matching
scale $Q_0^2$ consistent with QCD evolution equation~\cite{GRV9295,TVMZ97}.

We assume that the gluon distribution has a valence-like form and
reads
\be
G(x,Q_0^2) = \frac{N_g}{3} \left[u_V(x,Q_0^2) + d_V(x,Q_0^2)\right],
\label{eq:gluon}
\ee
where $N_g$ represents the number of gluons.
Since $\int dx\left[u_V(x,Q_0^2) + d_V(x,Q_0^2)\right] = 2+1$ because of baryon
number conservation, we have 
\be
\int dx\, G(x,Q_0^2) 
= \frac{N_g}{3} \int dx \left[u_V(x,Q_0^2) + d_V(x,Q_0^2)\right] = N_g.
\ee
Following Gl\"uck {\it et al.}~\cite{GRV9295}, 
we fix $N_g = 2$, the minimum number of gluons one needs to build a color
singlet state. 
 In spite of the simplicity of the assumption in eq.~(\ref{eq:gluon}),
the 
analysis of refs.~\cite{GRV9295,TVMZ97} shows that the crucial ingredient 
in the
formulation of the radiative parton model is the amount of gluon momentum more
than its actual form. 
Of course, to be really predictive one has to fit the form
of the gluon distribution in a quite precise way, an accuracy which goes beyond
the aim of the present study.

The total momentum carried at the scale $Q_0^2$ (the scale of the
{\it physical} proton, which will be larger than the scale of the {\it bare}
proton $\mu_0^2$ made up of three valence only) must fulfill the requirement
\be
\int dx\, x\left[
G(x,Q_0^2)+ u_V(x,Q_0^2) + d_V(x,Q_0^2)
+2\,\left[\bar u(x,Q_0^2) + \bar d(x,Q_0^2) \right] 
\right] = 1 ,
\ee
and consequently
\bea
&&\int dx\,x \left[u_V(x,Q_0^2) + d_V(x,Q_0^2)\right] \nonumber \\
&&\quad = \frac{1} {1+ \frac {N_g}{3}} 
\left[1 - \int dx\, 2\,x\left[\bar u(x,Q_0^2) + \bar d(x,Q_0^2) \right]\right] 
= 0.52 .
\eea
In conclusion, by using the notation $\langle f \rangle_n = \int dx\,x^{n-1}
f(x)$  for the moments, we have
$\langle u_V + d_V\rangle_2 = 0.52$, $\langle G \rangle_2 = 0.35$, 
$\langle 2\left(\bar u + \bar d \right)\rangle_2 = 0.13$,
consistent with a next-to-leading order (NLO) evolution of the moments (in the
DIS renormalization scheme) with $Q_0^2 = 0.27$ GeV$^2$ and
$\Lambda_{\rm NLO}= 0.248$ GeV~\cite{GRV9295,TVMZ97}.

The actual values of $Q_0^2$ are fixed~\cite{TVMZ97} evolving back unpolarized
data, until the valence distribution matches the required momentum
$\langle u_V + d_V\rangle_2 = 0.52$.
The procedure makes use of the valence contribution only, and 
it does not depend on the renormalization scheme.
The value of $\Lambda_{\rm NLO}$ is suggested by the analysis of Gl\"uck {\em et
al.}~\cite{GRV9295}, the coupling $\left.\alpha_s(Q_0^2)\right|_{\rm NLO}$  
is obtained evolving back the
valence distribution as previously discussed, and $Q_0^2$ is found 
by solving numerically the NLO transcendental equation 
\begin{equation}
\ln {Q_0^2\over\Lambda_{\rm NLO}^2}-{4\,\pi\over\beta_0\,\alpha_s} + 
{\beta_1\over\beta_0^2}\,\ln\left[
{4\,\pi\over\beta_0\,\alpha_s} + {\beta_1\over\beta_0^2}\right] = 0\,,
\label{0:10}
\end{equation}
which assumes the more familiar expression
\begin{equation}
{\alpha_s(Q^2) \over 4\pi}={1 \over \beta_0\ln(Q^2/\Lambda_{\rm NLO}^2)}
\left(1-{\beta_0 \over \beta_0^2}\,{\ln\ln(Q^2/\Lambda_{\rm NLO}^2)\over
\ln(Q^2/\Lambda_{\rm NLO}^2)}\right)
\label{1:10}
\end{equation}
only in the limit $Q^2\gg\Lambda_{\rm NLO}^2$ 
(an interesting discussion about the effects of the approximation (\ref{1:10}) 
can be found in ref.~\cite{WM96}).

The hadronic scale, $\mu_0^2$, consistent with the presence of valence 
degrees of freedom only (as discussed in sect.~\ref{sect:LFCQM}), is much lower 
and consistent with the constituent quark mass, its actual value being
$\mu_0^2 = 0.094$ GeV$^2$. The NLO evolution of the unpolarized distributions is
performed following the solution of the renormalization group equation in terms
of moments, i.e. 
$\langle f(Q^2)\rangle_n = \int_0^1 dx\,f(x,Q^2)\,x^{n-1}$, and involves kernels
which have been computed up to next-to-leading order (NLO) in perturbation
theory~\cite{BeMu}. 
Since, in our case, the starting points for the evolution 
($\mu_0^2, Q_0^2$) are rather low, 
the form of the equations must guarantee complete symmetry for the evolution 
from $\mu_0^2$ to $Q^2 \gg \mu_0^2$ and {\em back~}
avoiding additional approximations associated with Taylor expansions and not
with the genuine perturbative QCD expansion~\cite{TVMZ97,WM96}. 

\begin{figure}[ht]
\begin{center}
\includegraphics[width=28 pc]{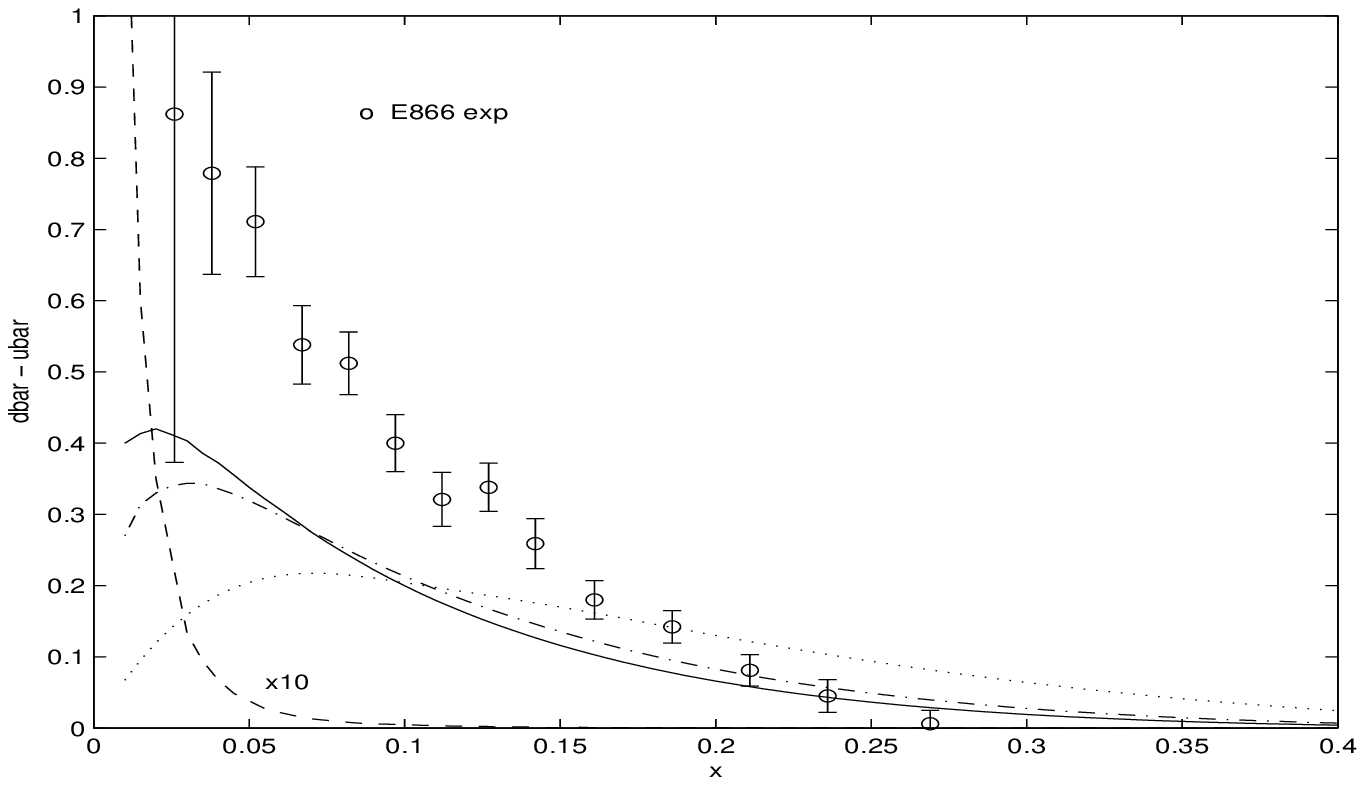}
\end{center}
\caption{$\bar d - \bar u$ distribution: at the scale $Q_0^2 = 0.27$
(GeV/c)$^2$ (dotted curve), at the (average) scale $Q^2 = 54$ (GeV/c)$^2$ of the
E866 experiment (solid curve), at the comparison scale of $Q^2 = 5$ (GeV/c)$^2$
(dot-dashed). The result (multiplied by a factor 10 for convenience) of an
evolution starting from the lowest hadronic scale where the nonperturbative 
sea content is neglected is also shown (dashed). Data from the E866 
experiment~\cite{E866}.}
\label{fig:fig1}
\end{figure}

In Fig.~\ref{fig:fig1} we compare our results for the $\bar d - \bar u$
difference with the data from the E866 experiment~\cite{E866}.
A QCD evolution with an SU(6) symmetric input introduces a very small
amount of asymmetric sea at the experimental scale. Within our approach 
such an assumption corresponds to the perturbative contribution coming from 
three valence quark distributions at the lowest hadronic 
scale as predicted by the relativistic model of
eq.~(\ref{eq:q-val})~\cite{BPT03}. The presence of the nonperturbative sea
introduced by means  
of the meson-cloud model discussed in the present section improves quite 
substantially the comparison with the experimental data. In particular
let us note: $i$) the important role played also by perturbative evolution
in the region $x < 0.1$ once the nonperturbative sea and gluon
content is introduced at the hadronic scale $Q_0^2$; $ii$) the satisfactory result
in spite of our simple model. All evolutions have been performed according to
ref.~\cite{TVMZ97}


\section{Results and discussion}
\label{sect:RES}

Results are presented in this section according to the model based on DDs as
described in sect.~\ref{sect:GPDs-DD}. In particular, we will take: $i$) forward
parton distributions with only valence quarks (section~\ref{sect:LFCQM}) at
input scale $\mu_0^2=0.094$ GeV$^2$ for the DDs, and $ii$) parton distributions
generated including meson-cloud corrections as in eq.~(\ref{eq:qp-cloud}) and
matched with the gluon distribution at an input scale $Q_0^2= 0.27$ GeV$^2$, as
described in section~\ref{sect:fromdiagonal}. 
In both cases our input GPDs are continuous functions all over the whole 
range $\overline x\in [-1,1]$. The $t$ dependence is dropped
from the very beginning and could be reintroduced in the final results by an
appropriate $t$-dependent factor. The D-term is omitted as well. 

In addition to the results at the hadronic scale we will discuss the evolution
of GPDs at higher scale.
The QCD evolution was numerically performed to NLO accuracy 
(in the $\overline{\mbox{MS}}$ scheme), according to a
modified version of the code of ref.~\cite{FmcD02b} (see
also~\cite{FmcD02a,FmcDS03}). The modifications we introduced in the main code
are basically due to the need of a complete 
NLO evolution which makes use of the correct NLO transcendental equation
(\ref{0:10}). The original code makes use of the simplified expression
(\ref{1:10}) largely unsatisfactory for our pourposes.

\begin{figure}[ht]
\begin{center}
\vspace{20pt}\epsfig{file=
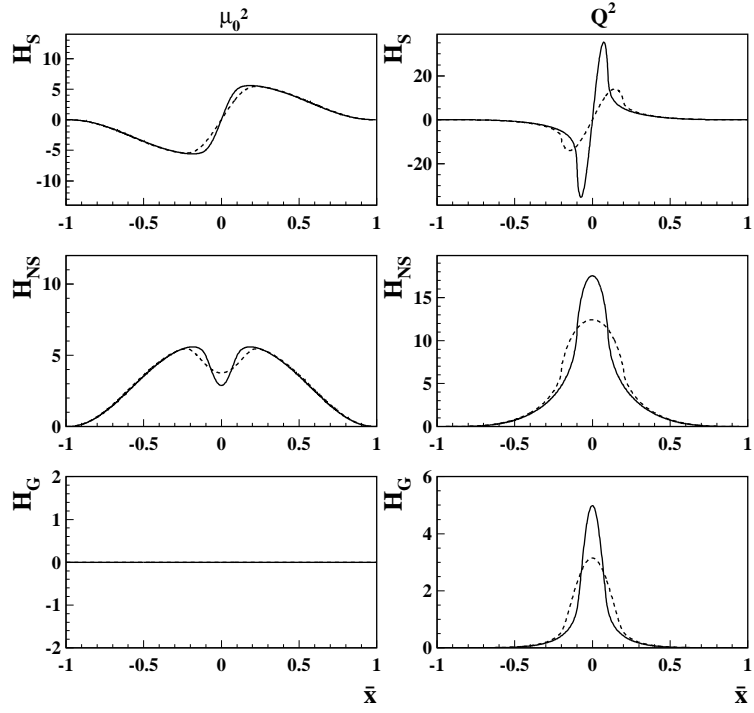, width=10cm}
\end{center}
\caption{Singlet quark (upper panels), non-singlet quark (middle panels), and
gluon (lower panels) GPDs using input parton
distributions of the hypercentral CQM. Full lines for $\xi=0.1$, dashed lines
for $\xi=0.2$. Left panels refer to the input hadronic scale $\mu_0^2$, 
right panels to the corresponding results obtained from NLO evolution
up to $Q^2=5$ GeV$^2$.}
\label{fig:sette}
\end{figure}

In fig.~\ref{fig:sette} results are shown for the singlet quark, non-singlet
quark and gluon GPDs obtained with no initial gluons and an input $q(\beta)$ in
eq.~(\ref{eq:ddunpol}) simply given by the bare (valence) parton distribution
$q_p^{bare}(x)$, eq.~(\ref{eq:q-val}), derived from the hypercentral CQM. The
model already gives a nonvanishing contribution to quark GPDs in the ERBL region
at the hadronic scale $\mu_0^2$ without introducing
discontinuities at $\overline x=\xi$ and with a weak $\xi$ dependence. 
In particular, the absence of the sea contribution gives $H^S=H^{NS}$ at
$\overline x>\xi$. After
evolution up to $Q^2=5$ GeV$^2$ GPDs are almost confined into the ERBL region
with a significant $\xi$ dependence.

This behavior is in agreement with previous studies of the GPD evolution showing
that as the resolution scale increases the distributions are swept from the
DGLAP domain to lie fully within the ERBL region~\cite{GoMa,shuvaev}. Functions
with support entirely in the time-like ERBL region $\vert \overline x\vert<\xi$
are never pushed out of the ERBL domain. In fact, the evolution in the ERBL
region depends on the DGLAP region, whereas the DGLAP evolution is independent
of the ERBL region. 

\begin{figure}[ht]
\begin{center}
\vspace{20pt}\epsfig{file=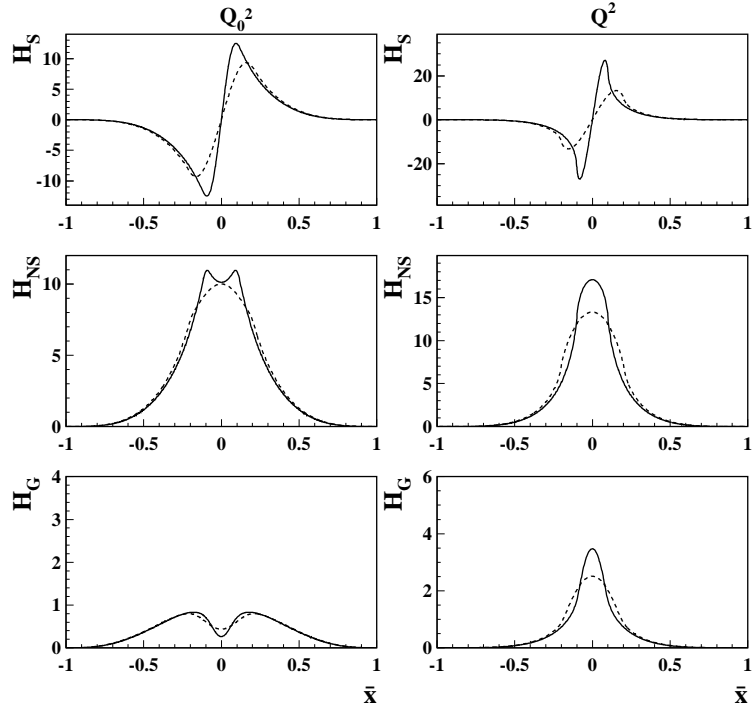, width=10cm}
\end{center}
\caption{Same as in fig.~\ref{fig:sette}, but with input parton distributions
including the sea contribution at the initial hadronic scale $Q_0^2$.}
\label{fig:otto}
\end{figure}

The same GPDs obtained when the input is implemented
with the sea according to the meson-cloud model, i.e. $q_p(x)$ of
eq.~(\ref{eq:qp-cloud}), are shown in fig.~\ref{fig:otto}. Now, the gluon GPD
does no more vanish at the initial hadronic scale, that has to be redefined at
$Q_0^2=0.27$ GeV$^2$ according to the conclusion of sect.~\ref{sect:Match}. 
After evolution
the qualitative result is similar to the case without the sea in spite of a more
evident $\xi$ dependence of the singlet quark GPD at the input hadronic scale.
However, the shape of the non-singlet quark GPD around $\overline x =0$ is
sensitive to the input, and the $\xi$ dependence of the GPDs size is in general
less pronounced.

\begin{figure}[ht]
\begin{center}
\vspace{20pt}\epsfig{file=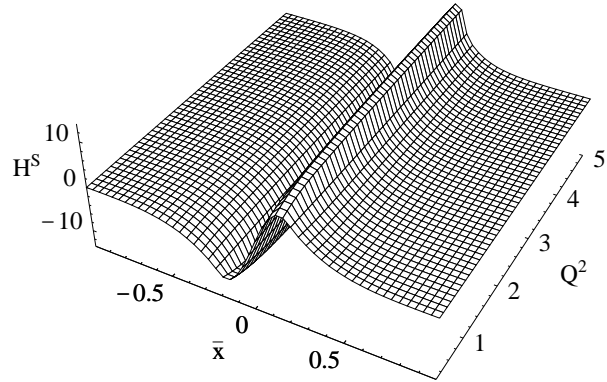, width=8cm}
\end{center}
\caption{Singlet quark GPD at $\xi=0.2$ as a function of $\overline x$ and
$Q^2$ obtained from NLO evolution of the GPDs using input parton distributions
of the hypercentral CQM with the sea 
contribution at the initial scale $Q_0^2$.} 
\label{fig:tred1}
\end{figure}

\begin{figure}[ht]
\begin{center}
\vspace{20pt}\epsfig{file=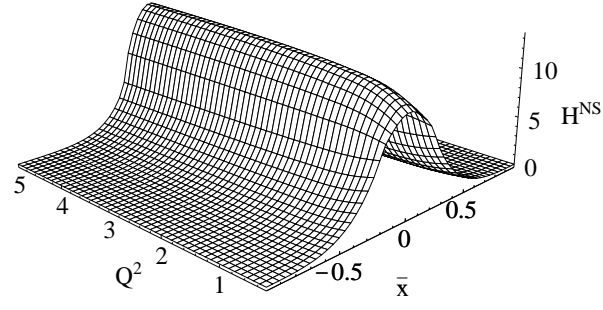, width=8cm}
\end{center}
\caption{Same as in fig.~\ref{fig:tred1}, but for the non-singlet quark GPD.} 
\label{fig:tred2}
\end{figure}

\begin{figure}[ht]
\begin{center}
\vspace{20pt}\epsfig{file=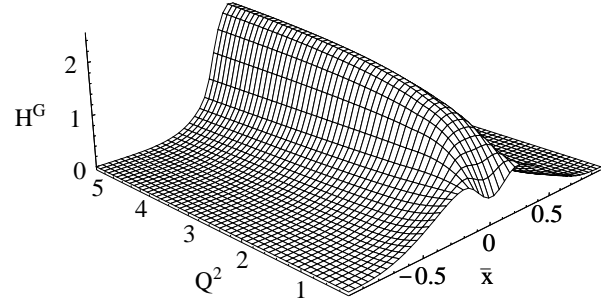, width=8cm}
\end{center}
\caption{Same as in fig.~\ref{fig:tred1}, but for the gluon GPD.} 
\label{fig:tred3}
\end{figure}

Results have been presented for QCD evolution up to $Q^2=5$ GeV$^2$. This is
already a value where GPDs have reached a stable configuration with respect to
their asymptotic shape. In fact, the largest effects of evolution modify the
input GPDs within the first few GeV$^2$ in the $Q^2$ evolution, as can be seen
in figs.~\ref{fig:tred1}-\ref{fig:tred3}, where the singlet quark, non-singlet
quark and gluon distributions are plotted at $\xi=0.2$ as a function of $Q^2$.
A similar behaviour has been also found for the valence $u$ GPD in the model of
Ref.~\cite{SV}.

\begin{figure}[ht]
\begin{center}
\vspace{20pt}\epsfig{file=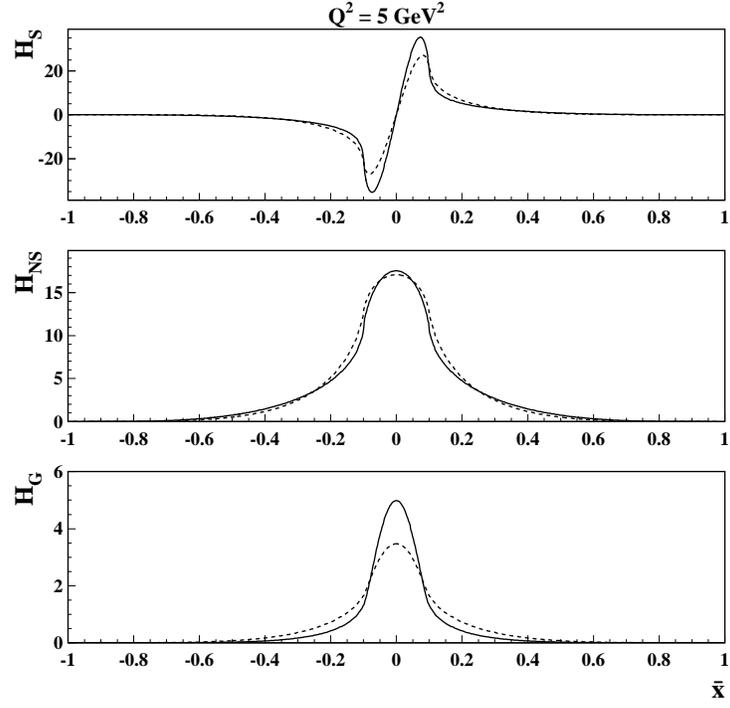, width=10cm}
\end{center}
\caption{Singlet quark (upper panels), non-singlet quark (middle panels), and
gluon (lower panels) GPDs at $\xi=0.1$ obtained from
NLO evolution up to $Q^2=5$ GeV$^2$. Full lines refer to input 
parton distributions
of the hypercentral CQM at the lowest hadronic scale $\mu_0^2=0.094$ GeV$^2$, 
dashed
lines refer to results including the sea contribution at the hadronic
scale $Q_0^2=0.27$ GeV$^2$.}
\label{fig:undici}
\end{figure}

The results discussed till now have been obtained with the usual choice 
$b_{val}=b_{sea}=1$ and $b_{gluon}=2$ (compare eq.~(\ref{eq:profile}) and 
the discussion of sect.~\ref{sect:GPDs-DD}). 
It is known~\cite{Musatov,Berger} that in the DD-based model
GPDs around $\overline x\sim\xi$ depend on forward densities at $\overline
x\ll\xi$, with a special sensitivity to the sea quark density. This is
particularly critical when using $q(x)$ and $xg(x)$ taken from the global parton
analyses~\cite{GRV98} because their singular behavior at $x=0$ is responsible
for a substantial and unrealistic enhancement of the quark singlet GPD relative
to parton density in the DGLAP region at $\overline x\sim\xi$. In the model
adopted here this singularity does not occur. Nevertheless, introducing the sea
has nonnegligible effects on the size of the ERBL response, especially for the
gluon distribution, as can be appreciated by looking at fig.~\ref{fig:undici},
where the same curves for $\xi=0.1$ from figs.~\ref{fig:sette} and
\ref{fig:otto} are directly compared.

\begin{figure}[ht]
\begin{center}
\vspace{20pt}\epsfig{file=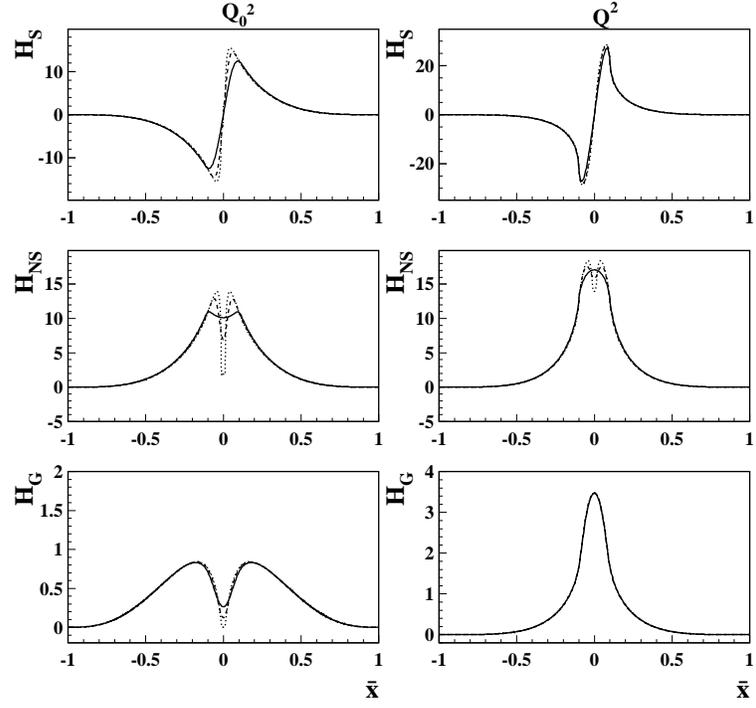, width=10cm}
\end{center}
\caption{Variation with the profile parameter $b$ of the singlet quark (upper
panels), non-singlet quark (middle panels), and gluon (lower panels) GPDs
derived for $\xi=0.1$ using input parton distributions of
the hypercentral CQM and including the sea contribution. Left panels refer to
the initial scale $Q^2_0$, right panels to the corresponding
results obtained from NLO evolution up to $Q^2=5$ GeV$^2$. 
Full lines for $b=1$,
dashed lines for $b=10$, and dotted lines for $b\to\infty$.}
\label{fig:buno}
\end{figure}

\begin{figure}[ht]
\begin{center}
\vspace{20pt}\epsfig{file=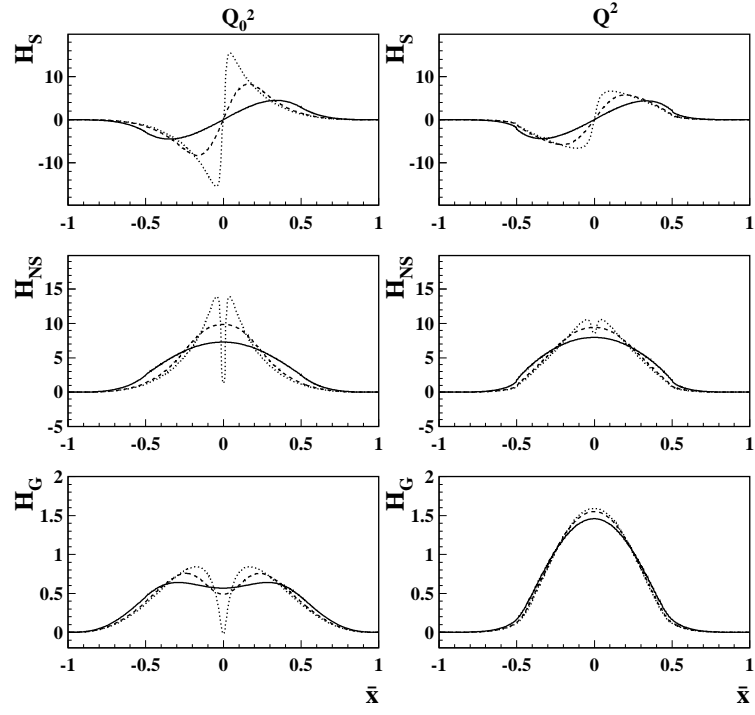, width=10cm}
\end{center}
\caption{Same as in fig.~\ref{fig:buno}, but for $\xi=0.5$.}
\label{fig:bdue}
\end{figure}

The value of the parameter $b$ in the profile function $h(\beta,\alpha)$,
eq.~(\ref{eq:profile}), determines its width and has been shown to be inversely
related to the enhancement of the singlet quark
GPD~\cite{Musatov,Berger,FmcDS03}. In principle, one could choose $b$ varying
from the adopted values up to infinity, with a Dirac-delta-like width and quark
GPDs reducing to parton densities. The $b$ dependence of GPDs in the DD-based
model including the sea contribution according to the meson-cloud model is
illustrated in figs.~\ref{fig:buno} and \ref{fig:bdue} for $\xi=0.1$ and
$\xi=0.5$, respectively. Only the ERBL region is affected by the choice of $b$.
At the input hadronic scale the $b$ dependence is strong, increases with $\xi$
and affects all GPDs, not only the non-singlet quark GPD. NLO evolution greatly
reduces such a $b$ dependence, and 
a similar behaviour is found for the evolution of GPDs also at LO.

\begin{figure}[ht]
\begin{center}
\epsfig{file=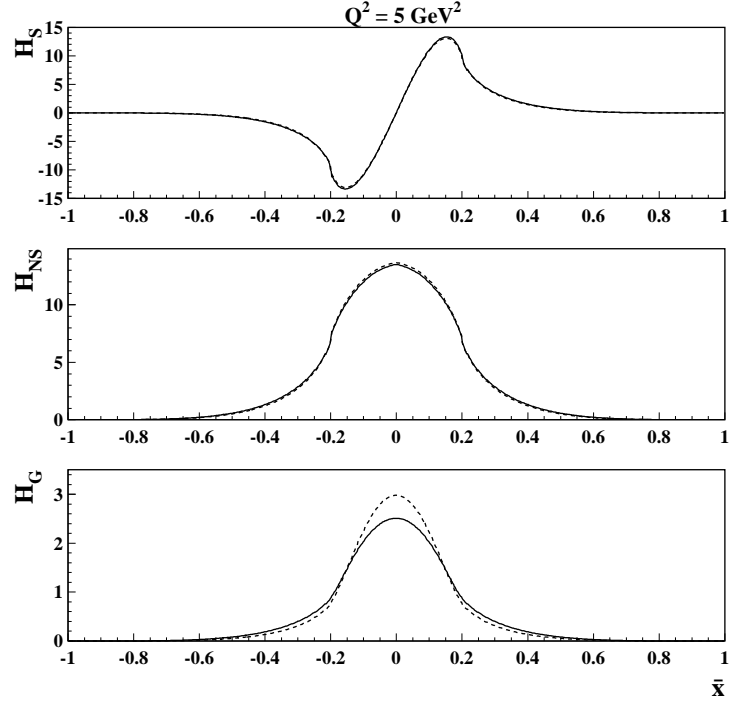, width=10 cm}
\end{center}
\caption{Singlet quark (upper panels), non-singlet quark (middle panels), and
gluon (lower panels) GPDs using input parton
distributions of the hypercentral CQM with the sea contribution in the 
meson-cloud model, evolved at 
$Q^2=5$ GeV$^2$ and at  $\xi=0.2$.
 The full lines are the results from NLO evolution, and the dashed lines 
corresponds to LO evolution.}
\label{fig:LONLO1}
\end{figure}

As a final remark let us discuss the validity of our evolution procedure.
As already mentioned the evolution has been performed by using a code due to
Freund and McDermott and modified to restore  the correct NLO coupling constant
thorough the transcendental equation (\ref{0:10}) (see discussion in the
previous section). The modification plays a crucial role when the evolution
starts from low hadronic input as in the present investigation.
With the caveat that perturbative stability can be tested only at the level
of physical cross sections or observables like the unpolarized structure 
function $F_2$ (see, for example, Ref.~\cite{patrabo}), 
we show in Fig.~\ref{fig:LONLO1} the results of LO and NLO evolution 
for singlet, nonsinglet and gluon distributions up to $Q^2=5$ GeV$^2$ 
starting from a hadronic scale as low as $Q_0^2=0.27$ GeV$^2$ (where
sea contributions are included). 
In the the singlet and nonsinglet sectors, the results at LO and at NLO
 are quite close, showing that the evolution is clearly under control 
and converging. 
Similar results hold also in the case of the evolution of GPDs 
from the lower scale of $\mu_0^2=0.094 $ GeV$^2$, without the sea contribution.
Only the gluon distribution shows larger discrepancy between LO and NLO 
evolution.
This is not a drawback of our model as can be seen
from analogous results of Ref.~\cite{FmcD02b}  which make use of
the standard (GRV98)~\cite{GRV98} parametrization for the diagonal 
partonic input at $Q_0^2 = 0.36$ GeV$^2$, 
comparable with our input scale.
In fact, it is well known that also in the case
of parton distribution functions generated from low scale parametrizations, NLO
gluons are affected by renormalization scale dependence, a problem not yet
addressed for GPDs and deserving further investigation.


\section {Conclusions}

Different inputs at the hadronic scale have been considered in the QCD evolution
of GPDs to study sensitivity of the results to the nonperturbative nature of the
low-scale hadronic structure. In particular, the meson-cloud model was assumed
to include sea quarks in the partonic content at the hadronic scale. Matching
the sea and gluon distributions with the valence-quark distributions derived in
a relativistic CQM, one starts evolution with continuous functions all over the
range $-1\le\overline x\le 1$. From the results presented here we can draw the
following conclusions. 

As already noticed in previous analyses~\cite{GoMa,shuvaev,GoMaRy}, evolution in
the DGLAP region is not much sensitive to the input. The reason is twofold. At
large $\overline x$ the valence contribution dominates at the input hadronic
scale and, as the scale increases, the distributions are swept from the DGLAP
domain to lie entirely in the ERBL region. In addition, evolution never pushes
the input distributions from the ERBL to the DGLAP region which then evolves
independently of the ERBL input. In contrast, the ERBL region is rather
sensitive to the input including or not the sea. 

Due to the mixing between singlet and gluon distributions in the evolution
equations, even with a vanishing gluon distribution at the hadronic scale one
obtains a significant gluon GPD in the ERBL region after evolution, peaked
around $\overline x=0$. The peak height depends on the input and is higher when
the sea is included.

The present results focus on the ERBL region as the most interesting one to look
at the nonperturbative effects surviving evolution. This is suggesting that one
has to investigate suitable processes under appropriate kinematic conditions to
study such effects. An analysis of possible observables is in progress.


\section*{Acknowledgements}

We are grateful to A. Freund and M. McDermott for providing us with their
evolution code. This research 
is part of the EU Integrated Infrastructure Initiative
Hadronphysics Project under contract number RII3-CT-2004-506078
and
was partially supported by the Italian MIUR
through the PRIN Theoretical Physics of the Nucleus and the Many-Body Systems.

 


\clearpage


\end{document}